\documentclass[onecolumn,showpacs,aps,epsfig,nofootinbib]{revtex4}

\usepackage{graphicx}
\usepackage{epstopdf}
\usepackage{latexsym}
\usepackage{amssymb}
\usepackage{amsmath}
\usepackage{color}
\usepackage{datetime}
\usepackage{extarrows}
\usepackage{mathrsfs,amssymb,bm}
\usepackage{hyperref}
\usepackage{multirow}
\usepackage{makecell}

\usepackage[center]{subfigure}

\begin{document}

 \newcommand{\bq}{\begin{equation}}
 \newcommand{\eq}{\end{equation}}
 \newcommand{\bqn}{\begin{eqnarray}}
 \newcommand{\eqn}{\end{eqnarray}}
 \newcommand{\nb}{\nonumber}
 \newcommand{\lb}{\label}
 \newcommand{\tc}{\textcolor{black}}
\newcommand{\PRL}{Phys. Rev. Lett.}
\newcommand{\PL}{Phys. Lett.}
\newcommand{\PR}{Phys. Rev.}
\newcommand{\CQG}{Class. Quantum Grav.}


\title{Preferred axis of CMB parity asymmetry in the masked maps}

\author{Cheng Cheng$^{1}$}

\author{Wen Zhao$^{2}$}
\email{wzhao7@ustc.edu.cn}

\author{Qing-Guo Huang$^{1}$}

\author{Larissa Santos$^{2}$}

\affiliation{ $^1$State Key Laboratory of Theoretical Physics, Institute of Theoretical Physics, Chinese Academy of Science, Beijing 100190, China \\ $^2$CAS Key Laboratory for Researches in Galaxies and Cosmology, Department of Astronomy, University of Science and Technology of China, Chinese Academy of Sciences, Hefei, Anhui 230026, China}

%


\begin{abstract}

Both WMAP and Planck data show a significant odd-multipole preference in the large scales of the cosmic microwave background (CMB) temperature anisotropies.  If this pattern originates from cosmological effects, then it can be considered a crucial clue for a violation in the cosmological principle. By defining various direction dependent statistics in the full-sky Planck 2015 maps (see, for instance, Naselsky et al. (2012); W. Zhao (2014)), we found that the CMB parity asymmetry has a preferred direction, which is independent of the choices of the statistics. In particular, this preferred axis is strongly aligned with those in the CMB quadrupole and octopole, as well as that in the CMB kinematic dipole, which hints to their non-cosmological origin. In realistic observations, the foreground residuals are inevitable, and should be properly masked out in order to avoid possible misinterpretation of the results. In this paper, we extend our previous analyses to the masked Planck 2015 data. By defining a similar direction dependent statistic in the masked map, we find a preferred direction of the CMB parity asymmetry, in which the axis also coincides with that found in the full-sky analysis. Therefore, our conclusions on the CMB parity violation and its directional properties are confirmed.

\end{abstract}


\maketitle

\section{Introduction}

The temperature and polarisation anisotropies of the cosmic microwave background (CMB) radiation, seeded by both primordial scalar and tensor fluctuations, encodes rich cosmological information \cite{cmb-review}. The precise measurements of the CMB power spectra by various experiments, including the WMAP and Planck satellites, have tightly constrained the cosmological parameters, showing an excellent consistency with the inflation+$\Lambda$CDM cosmological model \cite{wmap,planck}. At the same time, a number of anomalies on large scales were also reported in the CMB temperature anisotropy data \cite{wmap-anomaly,planck2013,planck2015,review}. The low quadrupole problem was first reported in COBE data \cite{cobe}, and later confirmed in WMAP, as well as, in Planck data. Other anomalies were also announced in both WMAP and Planck data, including the lack of both variance and correlation on the largest angular scales \cite{lack}, the alignment of the CMB quadrupole and octopole \cite{alignment}, the cold spot \cite{cold-spot}, the quadrant asymmetry \cite{quadrant}, the power asymmetry \cite{power}, and so on. All these anomalies seem to indicate an isotropy violation in the Universe on large scales. Different origins for the anomalies have been suggested throughout the years, from cosmological effects to unresolved contaminations or systematical errors, however, to the present time, they are still unknown.

The problem of the parity asymmetry of the CMB has been investigated in the literature \cite{parity1,kim2011}, and it shows a significant dominance of the power spectrum stored in the odd multipoles over the even ones. The odd parity preference was also confirmed in the recent Planck data \cite{planck2013,planck2015}. To distinguish between different explanations, we have investigated the directional properties of the CMB parity asymmetry by defining various direction-dependent statistics in previous works \cite{zhao2012,zhao2014}. We found that the CMB parity violation favors a preferred direction, which is independent of the choice of the statistics. Most importantly, we found that this preferred direction aligns with the direction of the CMB kinematic dipole. This coincidence strongly suggests that the parity asymmetry should be due to some unsolved systematical error related to the CMB dipole. In addition, by comparing the preferred directions in the parity asymmetry, and those in the CMB dipole, quadrupole and octopole, we found that the alignment between them is confirmed at more than $3\sigma$ confidence level. This shows that the CMB parity anomaly may have the same origin as the other anomalies: including the low quadrupole problem, the alignment between the quadrupole and the octopole, and the lack of large-scale correlation.

It is important to point out that in these previous works \cite{zhao2012,zhao2014}, we have used full-sky maps to construct the statistics for analysis. However, for any realistic CMB map, there are some foreground residuals. We previously assumed that these residuals have little effect on low multipoles, and cannot significantly influence our conclusion. However, the validity of this assumption is unknown, being, therefore, the main flaw of the previous analysis. The main goal of this paper is then to correct this flaw. In the present work, we shall mask the contaminated regions of the CMB maps, and investigate the directional properties of the CMB parity violation based on these masked maps. We find that the results using partial sky maps and full-sky maps are consistent with each other, which stabilizes our conclusions on the directional properties of CMB parity asymmetry.

\section{Preferred axis of CMB parity violation: Full CMB maps}

 According to coordinate transformation, the CMB temperature fluctuation on a two-dimensional sphere is a scalar field, which can be decomposed as the standard spherical harmonics as follows,
\begin{equation}
\Delta T(\hat{n})=\sum_{\ell=0}^{\infty} \sum_{m=-\ell}^{\ell} a_{\ell m} Y_{\ell m} (\hat{n}),
\end{equation}
where $Y_{\ell m}(\hat{n})$ are the spherical harmonics, and $a_{\ell m}$ are the corresponding coefficients. In the standard inflationary scenario, both primordial scalar and tensor perturbations are random Gaussian fields. In the linear order approximation, the two-dimensional temperature fluctuations also satisfy the random Gaussian distribution, i.e., the amplitudes $|a_{\ell m}|$ are distributed according to Rayleigh's probability distribution function, and the phase of $a_{\ell m}$ with $m\neq 0$ is supported to be evenly distributed in the range $[0,2\pi]$. The statistical properties of a random Gaussian field are completely described by the second-order power spectrum, namely
\begin{equation}
C_{\ell} \equiv \langle a_{\ell m}a^*_{\ell m}\rangle,
\end{equation}
where $\langle ... \rangle$ denotes the average over the statistical ensemble of realizations, and the spectrum $C_{\ell}$ is independent of the magnetic quantum number $m$ for the statistical isotropic field. In real detections, it is impossible to directly observe the power spectrum $C_{\ell}$. One has to construct the estimators. For the full-sky map, if any kind of noise is negligible, the best unbiased estimator for $C_{\ell}$ is \cite{grishchuk1997}
\begin{equation}\label{hat-cl}
\hat{C}_{\ell}=\frac{1}{2\ell+1} \sum_{m=-\ell}^{\ell} a_{\ell m} a_{\ell m}^*,
\end{equation}
and the statistical uncertainty is $\Delta \hat{C}_{\ell}=\sqrt{\frac{2}{2\ell+1}}C_{\ell}$, which is the so-called cosmic variance. Note that this estimator is rotationally invariant, i.e. its value is invariant under the rotation of the coordinate reference system. Based on the unbiased estimator $\hat{C}_{\ell}$, a statistic for the CMB parity asymmetry can be defined as
\begin{equation}\label{g}
G(\ell)=\frac{\sum_{\ell'=2}^{\ell}\ell'(\ell'+1)\hat{C}_{\ell'}\Gamma^{+}_{\ell'}}
{\sum_{\ell'=2}^{\ell}\ell'(\ell'+1)\hat{C}_{\ell'}\Gamma^{-}_{\ell'}},
\end{equation}
where $\Gamma^{+}_{\ell}=\cos^2(\ell \pi/2)$ and $\Gamma^{-}_{\ell}=\sin^2(\ell \pi/2)$. This statistic is associated with the degree of parity asymmetry, where a value $G<1$ indicates the odd-parity preference, and $G>1$ indicates the even-parity preference. In the full WMAP data, an odd-parity preference at the very large scales has been reported at a quite high confidence level. In the 7-year WMAP data, the minimum of $G$ in the lower-tail probability occurs at $\ell=22$, which corresponds to the probability value of $P=0.6\%$ \cite{kim2011}. While in the Planck 2015 data the minimum of $G$ extends to $\ell=28$, and the probability value also decreases to $0.2\%$ for NILC, SEVEM, and SMICA, and $0.3\%$ for Commander \cite{planck2015}. {{ Even so, as we have shown in a previous work \cite{zhao2012}, the main contribution to the CMB parity asymmetry comes from the lowest multipoles $\ell\le 10$. As a conservative discussion and in order to be consistent with previous works \cite{zhao2012,zhao2014}, we shall only consider the CMB low multipoles $\ell\le 21$. We expect the conclusions to be stable even if the multipoles up to $\ell \le 28$ are considered.}}

In this paper, we will study the directional properties of the CMB field as it has been done in previous works \cite{zhao2012,zhao2014}. The rotationally variant estimator $D(\ell)$ can be defined as follows
\begin{equation}\label{Dl}
\hat{D}_{\ell}=\frac{1}{2\ell} \sum_{m=-\ell}^{\ell} a_{\ell m} a_{\ell m}^*(1-\delta_{m0}),
\end{equation}
which is also an unbiased estimator for the power spectrum $C_{\ell}$, i.e., $\langle \hat{D}_{\ell} \rangle =C_{\ell}$. Comparing with $\hat{C}_{\ell}$ defined in Eq.(\ref{hat-cl}), the $m=0$ component has been excluded, so the $z$-axis in the referenced coordinate system has been selected as the preferred axis in this definition \cite{zhao2012,zhao2014}. Thus, we can construct the estimator $\hat{D}_{\ell}(\hat{\rm{\bf q}})$ in any coordinate system as
\begin{equation}\label{Dl-tilde}
\hat{D}_{\ell}(\hat{\rm{\bf q}})=\frac{1}{2\ell} \sum_{m=-\ell}^{\ell} a_{\ell m}(\hat{\rm{\bf q}}) a_{\ell m}^*(\hat{\rm{\bf q}})(1-\delta_{m0}),
\end{equation}
where we define that $\hat{\rm{\bf q}}\equiv (\theta,\phi)$, and $a_{\ell m}(\hat{\rm{\bf q}})$ are the coefficients in the coordinate system, which is the Galactic system rotated by the Euler angle $(0,\theta,\phi)$. Here $\hat{\rm{\bf q}}$ is a vector labeling the $z$-axis direction in the rotated coordinate system, and $(\theta,\phi)$ is the polar coordinate of this direction in the Galactic system. Similar to the previous works \cite{zhao2012,zhao2014}, the direction dependent statistic for the CMB parity asymmetry can be defined as
\begin{equation}\label{g}
g(\ell,\hat{\rm{\bf q}})=\frac{\sum_{\ell'=2}^{\ell}\ell'(\ell'+1)\hat{D}_{\ell'}(\hat{\rm{\bf q}})\Gamma^{+}_{\ell'}}
{\sum_{\ell'=2}^{\ell}\ell'(\ell'+1)\hat{D}_{\ell'}(\hat{\rm{\bf q}})\Gamma^{-}_{\ell'}}.
\end{equation}
Similar to $G(\ell)$, the statistic $g(\ell,\hat{\rm{\bf q}})$ is also associated with the degree of the parity asymmetry. For any given $\ell$, the sky map $g(\ell,\hat{\rm{\bf q}})$ can be constructed by considering all the directions $\hat{\rm{\bf q}}$. In practice, we pixelize the full sky in the HEALPix format with the resolution parameter $N_{\rm side}=64$ and set the directions $\hat{\rm{\bf q}}$ to be those of the pixels. From the definition of $g(\ell,\hat{\rm{\bf q}})$, the direction of $\hat{\rm{\bf q}}$ is exactly equivalent to its opposite direction $-\hat{\rm{\bf q}}$.

\begin{figure}[t]
\begin{center}
\includegraphics[width=17cm]{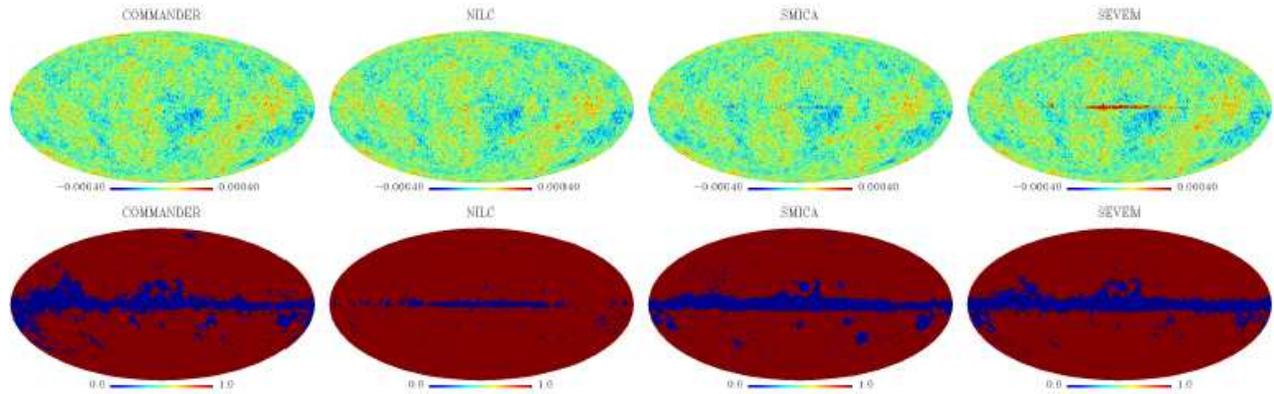}
\end{center}
\caption{The 2015 Planck temperature anisotropy maps, including Commander, NILC, SMICA and SEVEM. The lower panels are the corresponding masks suggested by Planck collaboration.}\label{fig0}
\end{figure}

\begin{figure}[t]
\begin{center}
\includegraphics[width=17cm]{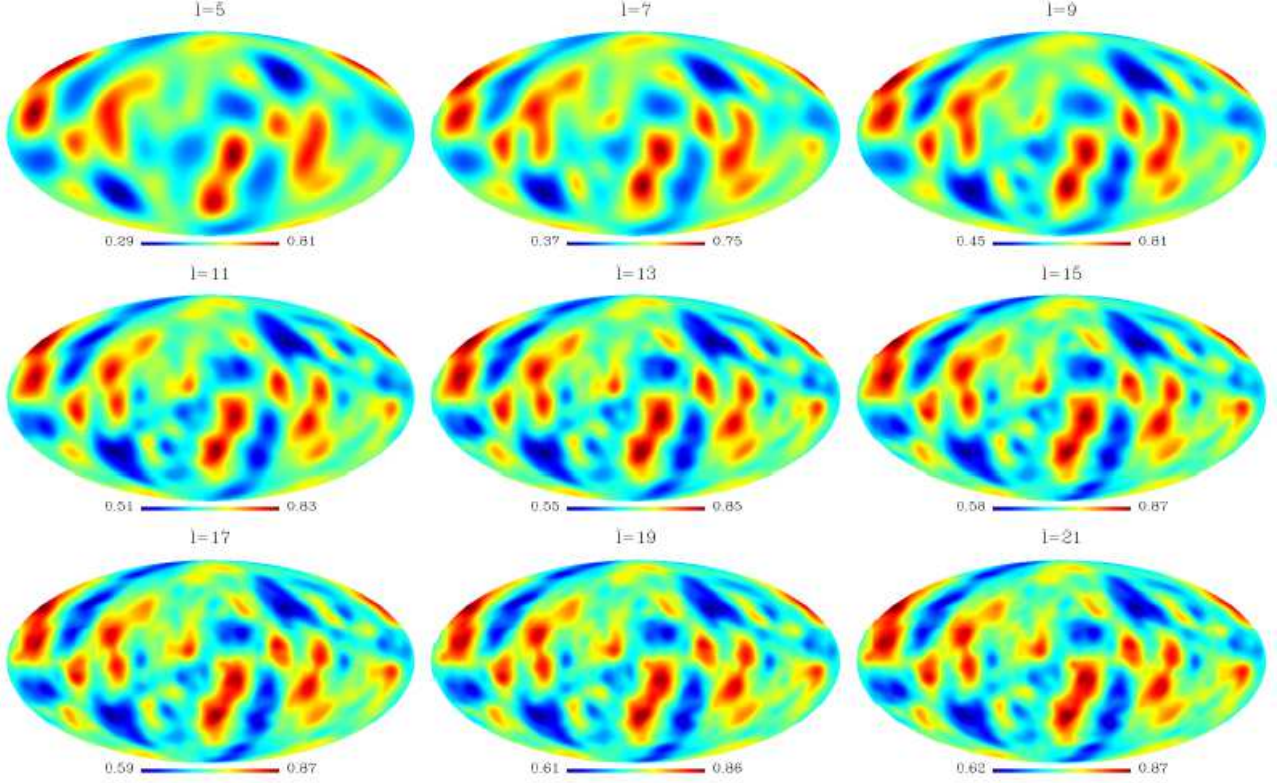}
\end{center}
\caption{The directional statistics $g(\ell,\hat{\rm{\bf q}})$ for different maximum multipole $\ell$ based on the full-sky Commander map.}\label{fig1}
\end{figure}

Recently, Planck collaboration released the 2015 data on the CMB temperature anisotropies, including Commander, NILC, SMICA and SEVEM. These CMB maps are constructed via quite different techniques, but they give the same cosmological results. From Fig. \ref{fig0}, we find that in the Galactic plane, these maps are definitely different due to the foreground residuals, and so providing different sky masks. However, similar to \cite{zhao2012,zhao2014}, we expect them to have little influence on the CMB low multipoles. For the consistency of our analysis, we shall consider all of these four maps. First of all, for example, let us consider the Commander map. Since the parity violation of the CMB is obvious only on large scales $\ell<22$ \cite{kim2011,planck2013,planck2015}, we only focus on the directional dependence of the CMB in this multipole range in this paper. Based on the definition of $g(\ell,\hat{\rm{\bf q}})$ in Eq. (\ref{g}), we construct $g$-maps for different maximum multipole $\ell$, which are presented in Fig. \ref{fig1}. From the figures, we find that $g(\ell,\hat{\rm{\bf q}})<1$ holds for any direction $\hat{\rm {\bf q}}$ and maximum multipole $\ell$, which is consistent with the discovery of the odd-parity preference in previous works \cite{kim2011,planck2013,planck2015}. The smaller the $g$-value the larger the parity violation. The figures show that, in the $g$-maps, the minimal $g$ value increases as the maximum multipole $\ell$ increases, which implies that the CMB parity violation mainly exists at the low multipole ranges. In addition, we find that for any given maximum multipole value $\ell$, except for the case with $\ell=3$, all the $g$-maps have quite similar morphologies. In all these maps, the preferred directions (listed in Table \ref{tab1}), where the $g$-value is minimized, are very close to each other. Note that, throughout this paper, we do not distinguish between the direction $\hat{\rm{\bf q}}$ and the opposite one $-\hat{\rm{\bf q}}$.

In previous works \cite{zhao2012,zhao2014}, we have found that the preferred axis coincides with the CMB kinematic dipole, which is independent of the choice of the directional statistics. In addition, the alignment of these preferred directions and the preferred directions in the CMB quadrupole and octopole is also confirmed at more than $3\sigma$ confidence level. Here, we also investigate this coincidence in the new data. To quantify the coincidence with the CMB kinematic dipole, we define the quantity $\alpha$, which is the angle between the preferred direction $\hat{\rm{\bf q}}$ and the CMB kinematic dipole direction at $(\theta=42^{\circ},\phi=264^{\circ})$ \cite{dipole}. We list the values of $|\cos\alpha|$ in Table \ref{tab1} and find that all of them are very close to 1, i.e. the angles $\alpha$ are all very close to zero. So, the coincidence with the CMB dipole is confirmed in the new released data.

{ The preferred directions in the CMB quadrupole and octopole are recently derived in the Planck data \cite{planck2013}, which slightly depend on the component-separated maps. For instance, for the SMICA map, they are $(\theta=13.4^{\circ},\phi=238.5^{\circ})$ for the quadrupole and $(\theta=25.7^{\circ},\phi=239.0^{\circ})$ for the octopole. While for the NILC map, they are $(\theta=12.7^{\circ},\phi=241.3^{\circ})$ for the quadrupole and $(\theta=25.8^{\circ},\phi=241.7^{\circ})$ for the octopole. The differences given by these two maps are quite tiny. So, in this paper, we shall only focus on the quadrupole and octopole derived from SMICA. In order to quantify the correlation between the preferred directions in the CMB parity asymmetry, the CMB dipole, the CMB quadrupole and the CMB octopole by a single statistic, we define the quantity (see, \cite{acceleration2,zhao2014}})
\begin{equation}
\langle|\cos\theta|\rangle=\sum_{i,j=1,j\neq i}^{N} \frac{|\hat{r}_i\cdot \hat{r}_j|}{N(N-1)},
\end{equation}
where $N$ is the number of directions. Here, for any given maximum multipole $\ell$, we consider four directions, namely the preferred direction $\hat{\rm{\bf q}}$, the CMB dipole direction, the CMB quadrupole preferred direction, and the CMB octopole preferred direction. In order to evaluate the significance of the alignment, we randomly generate $10^5$ samples, and in each sample four independent directions are randomly distributed in the two-dimensional sphere. From these samples, we obtain that $\langle|\cos\theta|\rangle=0.500\pm0.118$. Similar to previous works \cite{zhao2014,acceleration2}, we quantify the alignment by the ratio $\Delta_c/\sigma_c$, where $\Delta_c$ is the difference between the observed value of $\langle|\cos\theta|\rangle$ and the mean value of the simulations, and $\sigma_c$ is the corresponding standard deviation of the simulations. { According to the central limit theorem, the quantity $\langle|\cos\theta|\rangle$ satisfies the standard Gaussian distribution in the case of $N\rightarrow \infty$. Although, in our case, the value of $N$ is not very large, the quantity $\Delta_c/\sigma_c$ can also roughly evaluate the confidence level, as long as the observed result does not substantially deviate from the mean value of simulations}.
We list $\langle|\cos\theta|\rangle$ for all the $\ell$ cases in Table \ref{tab1}, and find that for any given maximum multipole $\ell$ the alignment is confirmed at more than $3\sigma$ confidence level, which is consistent with those in \cite{zhao2014}.

\begin{figure}[t]
\begin{center}
\includegraphics[width=17cm]{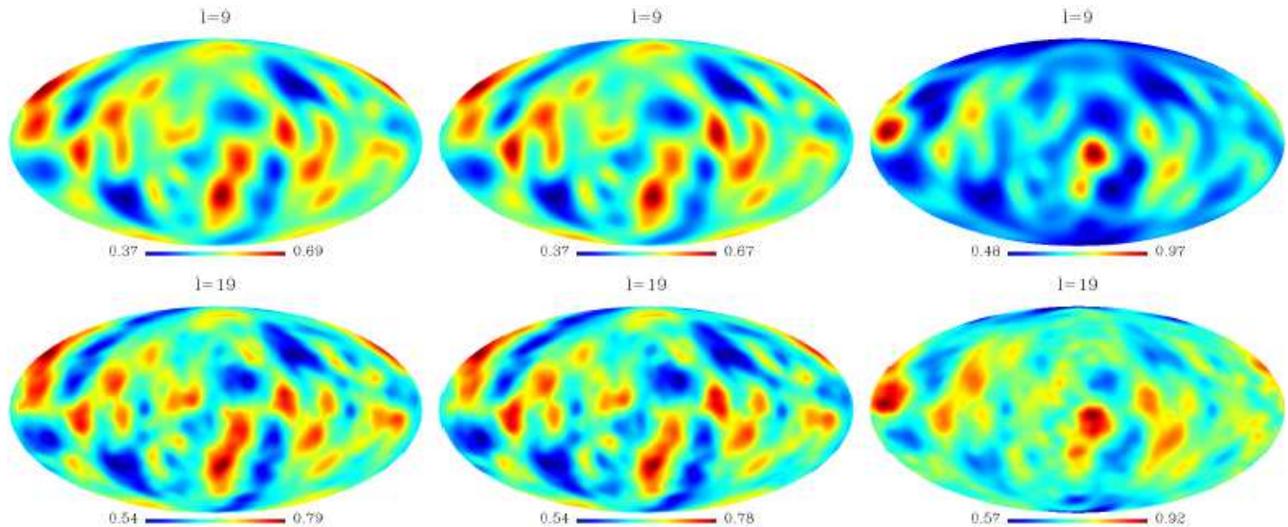}
\end{center}
\caption{The directional statistics $g(\ell,\hat{\rm{\bf q}})$ for different maximum multipoles $\ell=9$ (upper) and $\ell=19$ (lower) based on the full-sky NILC (left), SMICA (middle) and SEVEM (right) maps.}\label{fig2}
\end{figure}

Repeating the analysis for the NILC, SMICA and SEVEM maps, the corresponding results are all listed in Table \ref{tab2}-\ref{tab4}. We find that the Planck 2015 NILC and SMICA maps give almost the same results as those of Commander map. The alignment between the CMB parity asymmetry and the CMB dipole, quadrupole, and octopole are all confirmed. However, for the SEVEM map, the situation is slightly different. From Table \ref{tab4}, we find that although the ratios $\Delta_c/\sigma_c$ are still larger than 3, the preferred directions of the $g$-maps move to the ones close to the Galactic north pole, and the angle between $\hat{\rm{\bf q}}$ and the CMB kinematic dipole becomes a little bit larger, i.e., the coincidence with the dipole is weaker. This fact can be easily understood. From Fig. \ref{fig0}, we see that the Galactic region in the SEVEM map is quite dirty compared to the other three maps. So, we expect the effect of these foreground residuals in SEVEM map to be slightly larger than in the other maps even in the low multipoles. However, it is important to emphasize that from Fig. \ref{fig2} we can clearly see that even for the SEVEM map, the CMB dipole direction in the $g$-map is one of the preferred directions, even though it is not the most favoured one.

\begin{table}[!htb]
 \centering
 \begin{tabular}{c @{\extracolsep{4em}} c c c c c}
 \hline\hline
  &  & $\theta[^{\circ}]$  & $\phi[^{\circ}]$  &  $|\cos\alpha|$ & $\Delta_c/\sigma_c$\\
\hline
&$\ell=5$ & $45.03$ & $281.39$ & $0.977$ & $3.42$\\
        & & $45.82$ & $279.73$ & $0.980$ & $3.42$\\
        & & $45.03$ & $279.89$ & $0.981$ & $3.44$\\
\hline
&$\ell=7$ & $46.62$ & $279.58$ & $0.979$ & $3.40$\\
        & & $47.41$ & $278.00$ & $0.981$ & $3.39$\\
        & & $46.62$ & $278.12$ & $0.982$ & $3.41$\\
\hline
&$\ell=9$ & $47.41$ & $279.43$ & $0.978$ & $3.38$\\
        & & $52.86$ & $265.21$ & $0.977$ & $3.37$\\
        & & $47.41$ & $276.57$ & $0.984$ & $3.41$\\
\hline
&$\ell=11$ & $49.01$ & $278.58$ & $0.976$ & $3.34$\\
         & & $52.10$ & $267.32$ & $0.984$ & $3.35$\\
         & & $49.01$ & $274.36$ & $0.984$ & $3.38$\\
\hline
&$\ell=13$ & $49.01$ & $279.99$ & $0.973$ & $3.32$\\
        &  & $17.62$ & $129.44$ & $0.566$ & $2.91$\\
        &  & $49.01$ & $275.76$ & $0.982$ & $3.37$\\
\hline
&$\ell=15$ & $49.80$ & $283.50$ & $0.961$ & $3.26$\\
        &  & $19.85$ & $131.73$ & $0.546$ & $2.85$\\
        &  & $49.80$ & $280.69$ & $0.969$ & $3.29$\\
\hline
&$\ell=17$ & $50.57$ & $284.21$ & $0.957$ & $3.22$\\
        & & $19.85$ & $131.73$ & $0.546$ & $2.85$\\
        & & $50.57$ & $282.80$ & $0.961$ & $3.24$\\
\hline
&$\ell=19$ & $50.57$ & $284.21$ & $0.957$ & $3.22$\\
        & & $19.85$ & $135.07$ & $0.556$ & $2.88$\\
        & & $49.01$ & $270.14$ & $0.990$ & $3.42$\\
\hline
&$\ell=21$ & $50.57$ & $284.21$ & $0.957$ & $3.22$\\
        & & $20.59$ & $133.46$ & $0.543$ & $2.84$\\
        & & $50.57$ & $284.21$ & $0.957$ & $3.22$\\
\hline

\end{tabular}
\caption{The preferred direction $(\theta,\phi)$, and corresponding $|\cos\alpha|$ and $\Delta_c/\sigma_c$ for $g(\ell,\hat{\rm{\bf q}})$ based on Planck Commander map, where the different maximum multipole $\ell$ is considered. For each $\ell$ case, the upper values denote the results derived from the full-sky analysis, the middle values denote those derived from the masked case in which the Commander mask is applied, and the lower values denote those derived from the masked case in which NILC mask is applied. Note that the CMB kinematic dipole direction is at $(\theta=42^{\circ},\phi=264^{\circ})$, the preferred direction of CMB quadrupole is $(\theta=13.4^{\circ},\phi=238.5^{\circ})$ and that for the CMB octopole is at $(\theta=25.7^{\circ},\phi=239.0^{\circ})$.}
\label{tab1}
\end{table}

\begin{table}[!htb]
 \centering
 \begin{tabular}{c @{\extracolsep{4em}} c c c c c}
 \hline\hline
  &  & $\theta[^{\circ}]$  & $\phi[^{\circ}]$  &  $|\cos\alpha|$ & $\Delta_c/\sigma_c$\\
\hline
&$\ell=5$ & $45.82$ & $279.73$ & $0.980$ & $3.42$\\
        & & $45.82$ & $279.73$ & $0.980$ & $3.42$\\
\hline
&$\ell=7$ & $47.41$ & $278.00$ & $0.981$ & $3.39$\\
        & & $48.21$ & $275.06$ & $0.985$ & $3.40$\\
\hline
&$\ell=9$ & $48.21$ & $276.47$ & $0.982$ & $3.35$\\
        & & $49.80$ & $272.25$ & $0.985$ & $3.38$\\
\hline
&$\ell=11$ & $49.01$ & $277.17$ & $0.979$ & $3.35$\\
        & & $49.80$ & $272.25$ & $0.985$ & $3.38$\\
\hline
&$\ell=13$ & $49.01$ & $278.58$ & $0.976$ & $3.34$\\
        & & $49.80$ & $272.25$ & $0.985$ & $3.38$\\
\hline
&$\ell=15$ & $49.80$ & $282.10$ & $0.965$ & $3.27$\\
        & & $49.80$ & $272.25$ & $0.985$ & $3.38$\\
\hline
&$\ell=17$ & $50.57$ & $284.21$ & $0.957$ & $3.22$\\
        & & $49.80$ & $270.84$ & $0.987$ & $3.39$\\
\hline
&$\ell=19$ & $50.57$ & $284.21$ & $0.957$ & $3.22$\\
        & & $49.01$ & $270.14$ & $0.990$ & $3.42$\\
\hline
&$\ell=21$ & $50.57$ & $284.21$ & $0.957$ & $3.22$\\
        & & $49.01$ & $270.14$ & $0.990$ & $3.42$\\
\hline

\end{tabular}
\caption{The preferred direction $(\theta,\phi)$, and corresponding $|\cos\alpha|$ and $\Delta_c/\sigma_c$ for $g(\ell,\hat{\rm{\bf q}})$ based on Planck NILC map, where the different maximum multipole $\ell$ is considered. For each $\ell$ case, the upper values denote the results derived from the full-sky analysis, and the lower values denote those derived from the masked case in which NILC mask is applied.}
\label{tab2}
\end{table}

\begin{table}[!htb]
 \centering
 \begin{tabular}{c @{\extracolsep{4em}} c c c c c}
 \hline\hline
  &  & $\theta[^{\circ}]$  & $\phi[^{\circ}]$  &  $|\cos\alpha|$ & $\Delta_c/\sigma_c$\\
\hline
&$\ell=5$ & $45.03$ & $279.89$ & $0.981$ & $3.44$\\
        & & $45.03$ & $281.39$ & $0.977$ & $3.42$\\
        & & $45.82$ & $279.73$ & $0.980$ & $3.42$\\
\hline
&$\ell=7$ & $49.01$ & $277.17$ & $0.979$ & $3.35$\\
        & & $46.62$ & $279.58$ & $0.979$ & $3.40$\\
        & & $48.21$ & $275.06$ & $0.985$ & $3.40$\\
\hline
&$\ell=9$ & $49.80$ & $275.06$ & $0.981$ & $3.35$\\
        & & $53.60$ & $265.92$ & $0.979$ & $3.31$\\
        & & $50.57$ & $271.54$ & $0.984$ & $3.36$\\
\hline
&$\ell=11$ & $50.57$ & $277.17$ & $0.975$ & $3.31$\\
        & & $52.86$ & $266.62$ & $0.982$ & $3.33$\\
        & & $50.57$ & $271.54$ & $0.984$ & $3.36$\\
\hline
&$\ell=13$ & $49.80$ & $277.88$ & $0.976$ & $3.32$\\
        & & $16.14$ & $128.93$ & $0.582$ & $2.95$\\
        & & $51.34$ & $269.43$ & $0.984$ & $3.35$\\
\hline
&$\ell=15$ & $50.57$ & $281.39$ & $0.965$ & $3.26$\\
        & & $19.10$ & $129.87$ & $0.550$ & $2.86$\\
        & & $50.57$ & $271.54$ & $0.984$ & $3.36$\\
\hline
&$\ell=17$ & $50.57$ & $282.80$ & $0.961$ & $3.24$\\
        & & $19.85$ & $128.40$ & $0.537$ & $2.81$\\
        & & $50.57$ & $270.14$ & $0.986$ & $3.37$\\
\hline
&$\ell=19$ & $50.57$ & $284.21$ & $0.957$ & $3.22$\\
        & & $19.10$ & $133.34$ & $0.560$ & $2.89$\\
        & & $49.01$ & $270.14$ & $0.990$ & $3.42$\\
\hline
&$\ell=21$ & $50.57$ & $284.21$ & $0.957$ & $3.22$\\
        & & $20.59$ & $130.24$ & $0.533$ & $2.80$\\
        & & $49.80$ & $269.43$ & $0.988$ & $3.40$\\
\hline

\end{tabular}
\caption{Same to Table \ref{tab1}, but the Commander map and its correspondent mask are replaced by the SMICA map and the SMICA mask, respectively.}
\label{tab3}
\end{table}

\begin{table}[!htb]
 \centering
 \begin{tabular}{c @{\extracolsep{4em}} c c c c c}
 \hline\hline
  &  & $\theta[^{\circ}]$  & $\phi[^{\circ}]$  &  $|\cos\alpha|$ & $\Delta_c/\sigma_c$\\
\hline
&$\ell=5$ & $0.00$ & $179.36$ & $0.743$ & $3.38$\\
        & & $45.82$ & $279.73$ & $0.980$ & $3.42$\\
        & & $45.82$ & $279.73$ & $0.980$ & $3.42$\\
\hline
&$\ell=7$ & $0.00$ & $179.36$ & $0.743$ & $3.38$\\
        & & $48.21$ & $276.47$ & $0.982$ & $3.38$\\
        & & $48.21$ & $275.06$ & $0.985$ & $3.40$\\
\hline
&$\ell=9$ & $0.00$ & $179.36$ & $0.743$ & $3.38$\\
        & & $55.07$ & $263.10$ & $0.974$ & $3.27$\\
        & & $50.57$ & $270.14$ & $0.986$ & $3.37$\\
\hline
&$\ell=11$ & $0.00$ & $179.36$ & $0.743$ & $3.38$\\
        & & $55.07$ & $263.10$ & $0.974$ & $3.27$\\
        & & $50.57$ & $271.54$ & $0.984$ & $3.36$\\
\hline
&$\ell=13$ & $0.00$ & $179.36$ & $0.743$ & $3.38$\\
        & & $17.62$ & $129.44$ & $0.566$ & $2.91$\\
        & & $51.34$ & $269.43$ & $0.984$ & $3.35$\\
\hline
&$\ell=15$ & $4.39$ & $112.56$ & $0.696$ & $3.26$\\
        & & $19.85$ & $131.73$ & $0.546$ & $2.81$\\
        & & $19.10$ & $129.87$ & $0.550$ & $2.86$\\
\hline
&$\ell=17$ & $4.39$ & $112.56$ & $0.696$ & $3.26$\\
        & & $19.85$ & $128.40$ & $0.537$ & $2.81$\\
        & & $19.85$ & $128.40$ & $0.537$ & $2.81$ \\
\hline
&$\ell=19$ & $2.93$ & $101.30$ & $0.710$ & $3.29$\\
        & & $19.85$ & $135.07$ & $0.556$ & $2.88$\\
        & & $19.10$ & $129.87$ & $0.550$ & $2.86$\\
\hline
&$\ell=21$ & $0.00$ & $179.36$ & $0.743$ & $3.38$\\
        & & $20.59$ & $130.24$ & $0.533$ & $2.80$\\
        & & $20.59$ & $130.24$ & $0.533$ & $2.80$\\
\hline

\end{tabular}
\caption{Same to Table \ref{tab1}, but the Commander map and its correspondent mask are replaced by the SEVEM map and the SEVEM mask, respectively.}
\label{tab4}
\end{table}

\section{Preferred axis of CMB parity violation: Masked CMB maps}

In the CMB maps, various foreground residuals are always unavoidable, especially in the Galactic region. The foreground residuals for the CMB maps released by Planck in 2015 are shown in Fig. \ref{fig0}. Usually one anticipates that the effects of these residuals are small and negligible in the low multipole range. It is still worthy investigating the cases in which these contaminated data are excluded. The simplest way to exclude the polluted region is to apply the top-hat mask to the data. For each CMB map, the corresponding mask suggested by Planck collaboration is also shown in Fig. \ref{fig0} (lower panels). We find that the masked region in the Commander, SMICA and SEVEM maps are quite similar. While the masked region for the NILC map is quite small, and the information loss in the NILC map is expected to be much smaller than in the other three maps. For the masked map, the unbiased estimator for $C_{\ell}$ is not straightforward. A large number of methods have been suggested in the literature \cite{method1,method2,method3,method4,method5}. In this paper, we adopt the so-called pseudo-$C_{\ell}$ (PCL) estimator method \cite{method4}. Although PCL estimator is a suboptimal one, it can be easily realized in pixel space using fast spherical harmonics transformation, and has been applied to various CMB observations, including WMAP and Planck data.
Considering the window function $W(\hat{n})$, i.e. the mask, the pseudo coefficients $\tilde{a}_{\ell m}$ can be defined as
\begin{equation}
\tilde{a}_{\ell m}=\int \Delta T(\hat{n}) W(\hat{n}) Y_{\ell m}(\hat{n}),
\end{equation}
which is related to $a_{\ell m}$ by
\begin{equation}
\tilde{a}_{\ell m}= \sum_{\ell_{1}m_{1}} a_{\ell_{1}m_{1}} K_{\ell m\ell_{1}m_{1}}.
\end{equation}
The coupling matrix $K$ is given by
\begin{equation}
K_{\ell m\ell_{1}m_{1}}=\sqrt{\frac{(2\ell_1+1)(2l+1)}{4\pi}}\sum_{\ell_2 m_2} (-1)^{m}(2\ell_2+1) w_{\ell_{2}m_{2}}
\begin{pmatrix}
\ell_1 & \ell_2 & \ell \\
0 & 0 & 0
\end{pmatrix}
\begin{pmatrix}
\ell_1 & \ell_2 & \ell \\
m_1 & m_2 & -m
\end{pmatrix},
\end{equation}
and $w_{\ell m}$ are the coefficients of spherical harmonics expansion of the mask $W(\hat{n})$, i.e.,
\begin{equation}
w_{\ell m}=\int W(\hat{n}) Y_{\ell m}^* (\hat{n}) d\hat{n}.
\end{equation}

The pseudo estimator $\tilde{C}_{\ell}$ is defined analogously to (\ref{Dl}) in terms of the multipole coefficients $\tilde{a}_{\ell m}$ as
\begin{equation}
\tilde{C}_{\ell}=\frac{1}{2\ell+1} \sum_{m=-\ell}^{\ell} \tilde{a}_{\ell m} \tilde{a}^*_{\ell m}.
\end{equation}
The expectation value of $\tilde{C}_{\ell}$ is
$\langle \tilde{C}_{\ell} \rangle =\sum_{\ell'}C_{\ell'} M_{\ell \ell'}$,
where the coupling matrix is
\begin{equation}
M_{\ell \ell'}=(2\ell'+1)\sum_{\ell_{2}} \frac {2\ell_{2}+1} {4\pi}
\begin{pmatrix}
    \ell'&\ell_2&\ell \\
    0 & 0& 0
    \end{pmatrix}  ^{2}\tilde{w}_{\ell_{2}}
\end{equation}
and ${\tilde w}_{\ell}$ are the following power spectrum,
\begin{equation}
\tilde{w}_{\ell} = \frac{1}{2\ell+1}\sum_{m=-\ell}^{\ell}w_{\ell m}w^{*}_{\ell m}.
\end{equation}
Similarly the unbiased estimator in the masked sky can be constructed as
$\hat{\mathcal{C}}_{\ell}=\sum_{\ell'} M^{-1}_{\ell \ell'} \tilde{C}_{\ell'}$.
Note that, this unbiased estimator $\hat{\mathcal{C}}_{\ell}$ is also rotationally invariant. Actually, the general analyses of the CMB parity asymmetry are always based on the estimators of the CMB power spectrum in the masked space \cite{kim2011,planck2013,planck2015}.

In this paper, we focus on the direction dependence of the CMB parity violation. We then need the direction dependent estimators in advance. Similar to the Sec. II, we can build them by excluding the $m=0$ components for each multipole,
\begin{equation}\label{tilde-dl}
\tilde{D}_{\ell}=\frac{1}{2\ell} \sum_{m=-\ell}^{\ell} \tilde{a}_{\ell m} \tilde{a}^*_{\ell m}(1-\delta_{m0}).
\end{equation}
Which means, by definition, that the $z$-direction of the coordinate system is chosen as the preferred direction. However, the estimators $\tilde{D}_{\ell}$ are not unbiased. The expectation values are given by
$\langle \tilde{D}_{\ell} \rangle =\sum_{\ell'}C_{\ell'} N_{\ell \ell'}$,
where the coupling matrix $N_{\ell \ell'}$ is given by
\begin{equation}
N_{\ell \ell'}=M_{\ell \ell'}-\frac{2\ell' +1}{2\ell} \sum_{\ell_{2} \ell_{2}^{'}m_{1}} \frac {\sqrt{(2\ell_{2} +1)(2\ell_{2}^{'} +1)}} {4\pi}
\begin{pmatrix}
    \ell'&\ell_2&\ell \\
    0 & 0& 0
\end{pmatrix}
\begin{pmatrix}
    \ell'&\ell_2^{'}&\ell \\
    0 & 0& 0
\end{pmatrix}
\begin{pmatrix}
    \ell'&\ell_2&\ell \\
   m_1 & -m_1& 0
\end{pmatrix}
\begin{pmatrix}
    \ell'&\ell_2&\ell \\
    m_1 & -m_1& 0
\end{pmatrix} w_{\ell_2 m_1} w_{\ell_2^{'} m_1}.
\end{equation}
Based on this relation, we can construct the unbiased estimator $\hat{\mathcal{D}}_{\ell}$ as follows,
\begin{equation}
\hat{\mathcal{D}}_{\ell}=\sum_{\ell'} N^{-1}_{\ell \ell'} \tilde{D}_{\ell'}.
\end{equation}
Similar to $\hat{D}_{\ell}$, $\hat{\mathcal{D}}_{\ell}$ are also the coordinate dependent unbiased estimators for the power spectra $C_{\ell}$, and the preferred direction is also the $z$-direction of the corresponding coordinate system.

For any coordinate system, the direction-dependent unbiased estimator $\hat{\mathcal{D}}_{\ell}(\hat{\rm{\bf q}})$ can be  built in the same manner as $\hat{\mathcal{D}}_{\ell}$, but the coefficients $\tilde{a}_{\ell m}$ and ${w}_{\ell m}$ are replaced by $\tilde{a}_{\ell m}(\hat{\rm{\bf q}})$ and ${w}_{\ell m}(\hat{\rm{\bf q}})$. The direction-dependent statistic for the CMB parity asymmetry can be defined as
\begin{equation}
{g}(\ell,\hat{\rm{\bf q}})=\frac{\sum_{\ell'=2}^{\ell}\ell'(\ell'+1)\hat{\mathcal{D}}_{\ell'}(\hat{\rm{\bf q}})\Gamma^{+}_{\ell'}}
{\sum_{\ell'=2}^{\ell}\ell'(\ell'+1)\hat{\mathcal{D}}_{\ell'}(\hat{\rm{\bf q}})\Gamma^{-}_{\ell'}}.
\end{equation}
Since $\hat{\mathcal{D}}_{\ell}$ are the unbiased estimators for the power spectra $C_{\ell}$, the new statistic $g(\ell,\hat{\rm{\bf q}})$ also indicates the degree of the CMB parity asymmetry and its direction dependence. Comparing with the ideal case using full-sky map and negligible noise, applying the mask on the data affects the values of the statistic $g$ in two aspects: 1) the CMB information is lost in the masked region, and the values of the unbiased estimators for the power spectrum $C_{\ell}$ and their uncertainties might be influenced; 2) the structure and position of the mask may influence the preferred direction of the $g$-maps by the definition of the directional estimator $\tilde{D}_{\ell}$ in Eq. (\ref{tilde-dl}). If the masked region is small, we expect both effects to be negligible, and the results for the masked case should be very close to the ideal case.

\begin{figure}[t]
\begin{center}
\includegraphics[width=17cm]{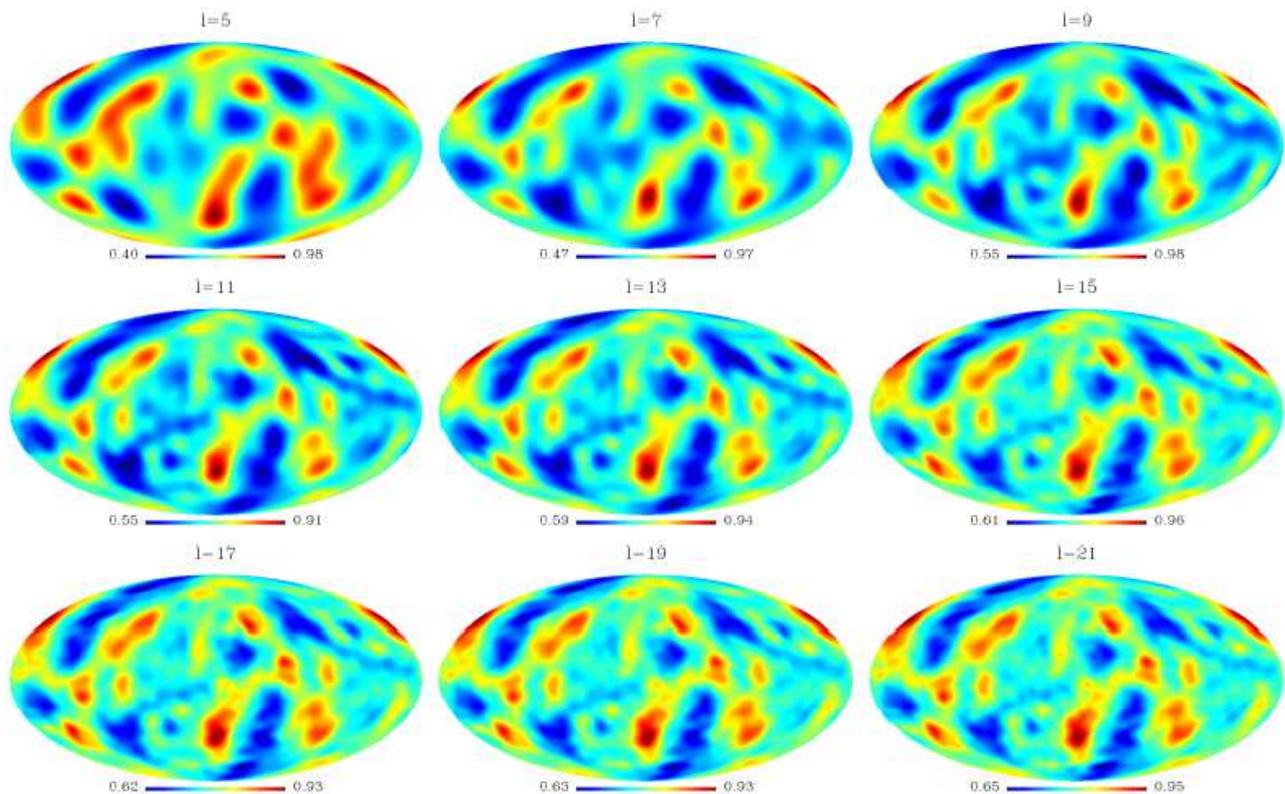}
\end{center}
\caption{The directional statistics $g(\ell,\hat{\rm{\bf q}})$ for different maximum multipole $\ell$. Note that, here we have considered the masked Commander map by applying Commander mask.}\label{fig3}
\end{figure}

\begin{figure}[t]
\begin{center}
\includegraphics[width=17cm]{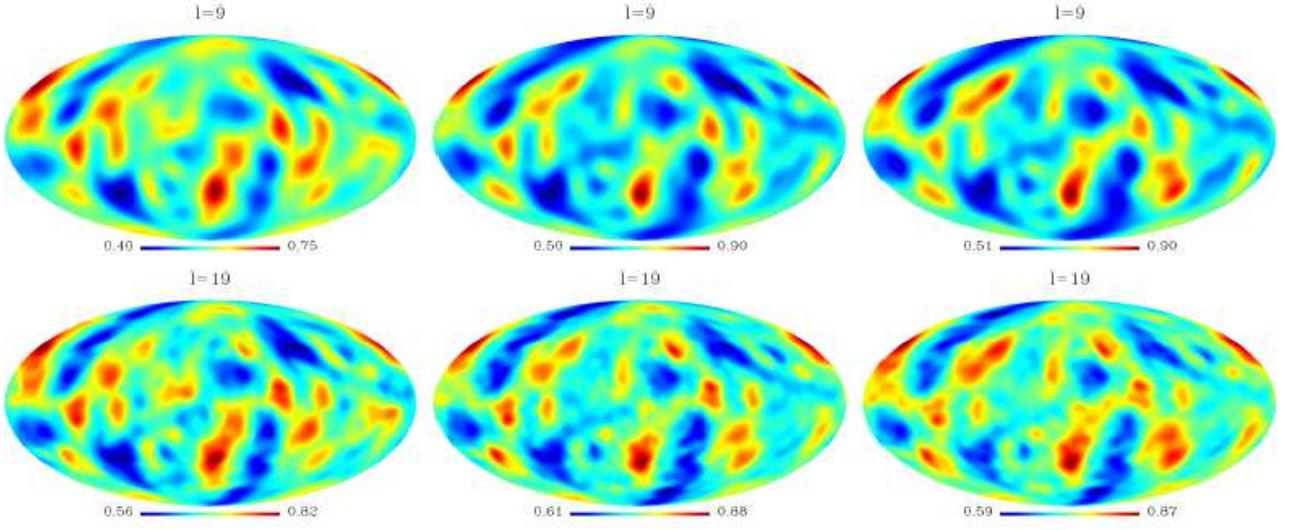}
\end{center}
\caption{The directional statistics $g(\ell,\hat{\rm{\bf q}})$ for different maximum multipoles $\ell=9$ (upper) and $\ell=19$ (lower) based on the masked NILC (left, by applying NILC mask), SMICA (middle, by applying SMICA mask) and SEVEM (right, by applying SEVEM mask) maps.}\label{fig4}
\end{figure}

Based on the estimators in the masked maps, we plot the $g$-maps for different maximum multipole $\ell$ and different CMB maps in Fig. \ref{fig3} and Fig. \ref{fig4}. Comparing with Figs. \ref{fig1} and \ref{fig2}, we find that the morphological structures of the $g$-maps slightly change due to the mask effect, but the cold and hot regions remain. For each map and each multipole case, the CMB dipole direction is one of the preferred directions of the $g$-map (although it may not be the most favoured one). In Tables \ref{tab1}-\ref{tab4}, we also list the most favoured directions $(\theta,\phi)$, and the values of $|\cos\alpha|$ and $\Delta_c/\sigma_c$ for the masked cases. First of all, let us focus on the NILC map, in which the mask region is very small (see Fig. \ref{fig0}). Comparing the left panels in Fig. \ref{fig4} with those in Fig. \ref{fig2}, as anticipated, we find the corresponding $g$-maps nearly the same. From Table \ref{tab2}, we find that the preferred directions in the masked case are very close to those in the full-sky case. The alignment between the preferred direction, the CMB kinematic dipole direction, the CMB quadrupole preferred direction and the CMB octopole preferred direction is confirmed at more than $3\sigma$ confidence level. In addition, $|\cos\alpha|>0.98$ are held for all maximum multipole cases, which means that the angle between the preferred direction $\hat{\rm{\bf q}}$ and the CMB dipole direction are all smaller than $11.5^{\circ}$.

\begin{figure}[t]
\begin{center}
\includegraphics[width=17cm]{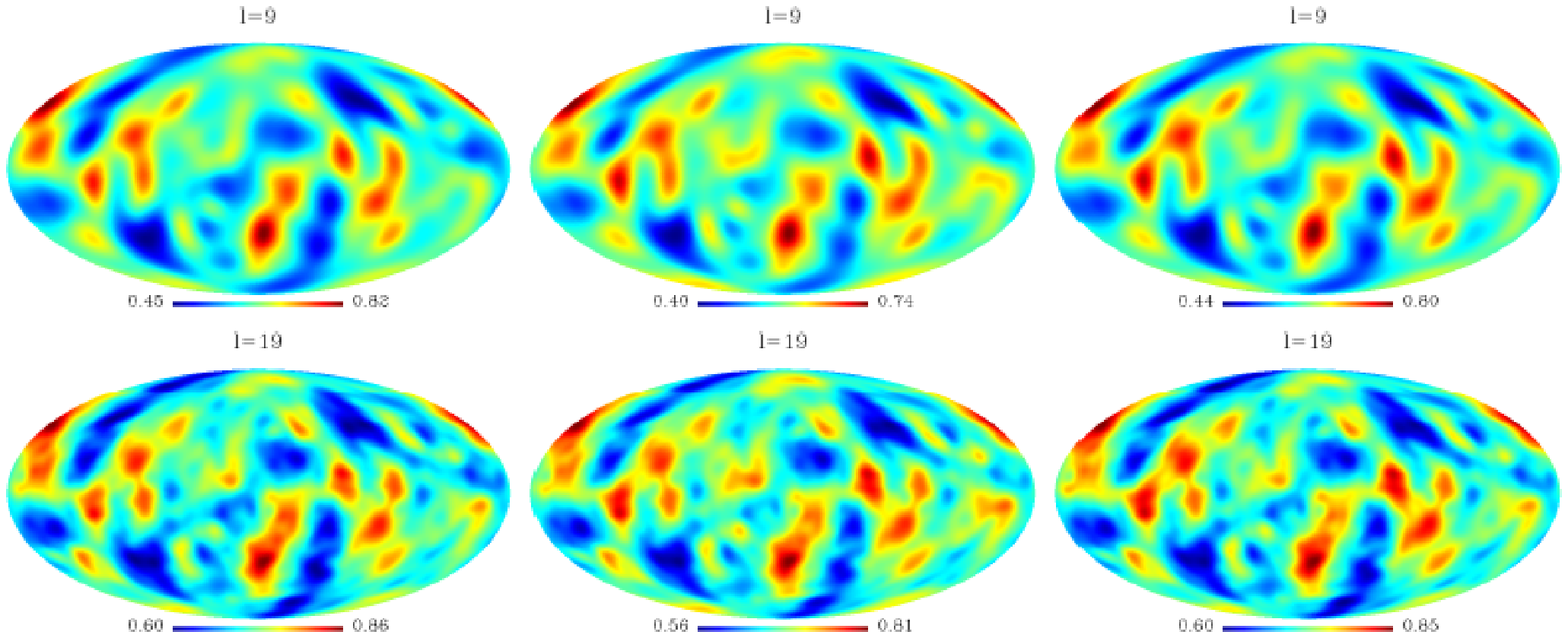}
\end{center}
\caption{The directional statistics $g(\ell,\hat{\rm{\bf q}})$ for different maximum multipoles $\ell=9$ (upper) and $\ell=19$ (lower) based on the masked Commander (left), SMICA (middle) and SEVEM (right) maps. Note that, in the figure, we have applied the NILC mask to all the three maps.}\label{fig5}
\end{figure}

For Commander and SMICA maps, from Fig. \ref{fig0}, we know that the masked regions are quite large, and they are mainly in the Galactic plane. From Figs. \ref{fig3} and \ref{fig4},  we find that in the low multipole range $\ell_{\rm max}\le 11$, the distributions of the $g$-maps based on the masked maps are quite similar to the those derived from the full-sky maps. So, the preferred directions in these $g$-maps strongly coincide with the preferred directions of the CMB dipole, quadrupole, and octopole (see Tables \ref{tab1} and \ref{tab3}). These results can be easily understood: even for the masked map, the original low multipoles can be nearly reconstructed \cite{ma2010}, being, therefore, the $g$-maps built from the masked CMB maps very similar to those built from the ideal full-sky maps. However, in the higher multipoles, the effects of the masks become larger, and the distributions of $g$-maps and the most favoured directions slightly change (see the results presented in Figs. \ref{fig3} and \ref{fig4} as well as in Tables \ref{tab1} and \ref{tab3}). However, the results change for the masked Commander and SMICA maps in the cases of $\ell_{\rm max}\geq 13$. This deviation might be caused by the mask application. We notice that the contaminations of Commander and SMICA are quite small and mainly concentrate on the thin band of the Galactic plane, and we suppose that most of the contaminations of these two maps can be nearly removed even if the NILC mask is applied (see Fig. \ref{fig0}). We repeat the analysis for the masked Commander and SMICA maps, but the masks are replaced by the NILC mask. Since the masked region is quite small and the unmasked region is clean enough, the influences of both foreground residuals and applied mask are small, being the results close to the real physics. The results are presented in Fig. \ref{fig5} and Tables \ref{tab1}-\ref{tab4}. We see that the preferred axes are very close to the full-sky cases even for $13\le \ell_{\rm max}\le 21$. So, we conclude that for any $\ell\le 21$ case the preferred axis is aligned with those in the CMB quadrupole and octopole, as well as with that of the CMB kinematic dipole.


Now, let us turn ourselves to the SEVEM map. In Sec. II, since the full-sky SEVEM map is very dirty, the results from the SEVEM full-sky map are not reliable. By applying the SEVEM mask (the SEVEM mask is similar to the Commander mask or the SMICA mask), the contaminations can be well avoided, but the effect of masking becomes large because a too large region is removed. From Fig. \ref{fig3} and Table \ref{tab4}, we find that the results in this case are the same as those in the case where Commander map is masked by Commander mask, or those in the case where SMICA map is masked by SMICA mask. Even so, we also find that if the maximum multipole is $\ell\le 11$, all analyses based on the masked SEVEM map (applying NILC mask and applying SEVEM mask) give similar results, which shows that the effects of both contamination and masks on the lowest multipoles are negligible.


\section{Discussions and conclusions}

The precise observations on the CMB temperature anisotropies provide an excellent way to test the isotropy of the Universe on the largest scales. Recent observations by the WMAP and Planck satellites indicated a number of anomalies, mainly on large scales, which may hint to the violation of the cosmological principle. Among them, the CMB parity asymmetry in the low multipoles $\ell\lesssim 30$ has been confirmed by Planck 2013 and 2015 data. In previous works \cite{zhao2012,zhao2014}, by defining several different directional statistics, we found that the CMB parity asymmetry has a preferred direction which is independent of the choice of the maximum multipole $\ell$ or the definition of the statistic. In particular, we found that this preferred direction is strongly aligned with the direction of the CMB kinematic dipole and the preferred directions of the CMB quadrupole and octopole. { The correlation of the preferred direction in the CMB parity asymmetry and those in the CMB quadrupole and octopole hints that the parity asymmetry in the CMB is not an isolated anomaly and it should have an intrinsic relationship with other anomalies, including the low multipole anomaly, the alignment of the CMB quadrupole and octopole, and the lack of large-scale correlations.}

In all of the previous works \cite{zhao2012,zhao2014}, we have considered the full-sky WMAP and Planck maps, and assumed that the effects of foreground residuals around the Galactic plane on the low multipoles are small enough. In this paper, we repeated the analysis for the Planck 2015 data (i.e. Commander, NILC, SMICA and SEVEM maps), and obtained the same results as before. In particular, as a consistency check, we considered the CMB masks which are applied to remove the influence of various contaminations. For the masked maps, we have used the pseudo-$C_{\ell}$ method to construct the unbiased but direction-dependent estimators of the CMB power spectrum, and the corresponding directional statistics for the CMB parity asymmetry. We found that if the masked region is small (e.g. in the NILC case), the results of the direction dependence in the CMB parity asymmetry derived from the masked maps are the same as those derived from the full-sky map, which stabilizes our conclusions. For the Commander, SMICA and SEVEM maps, in which the masked regions are quite large, we still found consistent results with the full-sky analysis in the low multipole range $\ell\le 11$.

Actually, the anisotropy problems have been reported not only on the CMB low multipoles, but also in a number of other cosmological observations: including the velocity flows \cite{velocity}, quasar alignment \cite{quasar}, anisotropies of cosmic acceleration \cite{acceleration}, the handedness of spiral galaxies \cite{spiral}, and angular distribution of the fine structure constant \cite{fine}. Even though there still are many debates in the literature \cite{debate1,debate2,debate3,debate4,debate5}, all these preferred directions seems to coincide with the CMB kinematic dipole.
These coincidences might imply the same origin for these anomalies:\\
1) A special topology of the Universe (for instance, the Bianchi model is suggested to replace the Friedmann-Robertson-Walker model in order to describe the metric of the Universe \cite{planck2013});\\
2) A special theory of gravity (for instance, in \cite{finsler}, the authors suggested to use the Finsler Gravity to replace General Relativity, or in \cite{deSitter}, the de Sitter Relativity is suggested to replace the General Relativity); \\
3) Some foreground residuals in the Solar System \cite{solar} or in the Galaxy \cite{galactic}. \\
However, in all these models, it is difficult to explain why the preferred direction in cosmology or gravitational physics coincides with the direction of the CMB kinematic dipole, which has been confirmed to be caused mainly by the motion of our local group of galaxies relative to the reference frame of the CMB. So, we believe that these cosmological anomalies should be caused by some unknown dipole-related systematics or contamination. For instance, in \cite{liu}, the authors found that the CMB kinematic dipole deviation could generate the artificial CMB anisotropies on low multipoles. If this is true, these artificial components may account for some direction-dependent CMB anomalies. Another possibility is that the preferred direction is caused by the tidal field originated from the anisotropy of our local halo. In \cite{wang}, the authors found that the tidal field tends to preferentially align with the orientation and spatial distribution of galaxies, which may also generate some unsolved kinematic or higher order effects, and influence the cosmological observations \cite{wang2}. We expect that future measurements on the CMB polarisation, cosmic weak lensing, or distribution of 21-cm line can help us to solve the puzzles.

\section*{Acknowledgements}
We acknowledge the use of the Planck Legacy Archive. Our data analysis made the use of HEALPix
\cite{healpix}.
W.Z. is supported by Project 973 under Grant No. 2012CB821804, by NSFC No. 11173021, 11322324, 11421303 and project of KIP and CAS. Q.G.H. is supported by Top-Notch Young Talents Program of China and grants from NSFC (grant NO. 11322545, 11335012 and 11575271). Q.G.H. would also like to thank the participants of the advanced workshop ``Dark Energy and Fundamental Theory" supported by the Special Fund for Theoretical Physics from the National Natural Science Foundations of China (grant No. 11447613) for useful conversation.

\end{document}